\documentclass[sigconf,nonacm]{acmart}

\usepackage{booktabs}
\usepackage{graphicx}
\graphicspath{{./}}
\usepackage{amsmath}
\usepackage{listings}
\usepackage{placeins} 

\begin{document}

\title{Auditing Retrieval-Augmented LLM Hypotheses for Longitudinal Cell Painting Morphology}

\author{Gilchan Park, Guang Zhao, Byung-Jun Yoon, Shinjae Yoo}
\affiliation{%
  \institution{Brookhaven National Laboratory}
  \city{Upton, New York}
  \country{USA}
}
\email{{gpark, gzhao, byoon, sjyoo}@bnl.gov}

\acmConference[ACM-BCB '26]{ACM Conference on Bioinformatics, Computational Biology, and Health Informatics}
\acmBooktitle{ACM Conference on Bioinformatics, Computational Biology, and Health Informatics (ACM-BCB '26), June 30--July 3, Rende (Cosenza), Italy}

\begin{abstract}
High-content morphological profiling (Cell Painting) yields sensitive, high-dimensional signatures of cellular state, but translating longitudinal morphology trajectories into interpretable biology remains difficult—especially for weak, chronic perturbations such as low-dose-rate ionizing radiation. Large language models (LLMs) can synthesize heterogeneous evidence into narratives, yet scientific use requires quantitative auditing: fluent outputs may be ungrounded or inconsistent with measured morphology.

We present an evaluation-first, retrieval-augmented interpretation framework for longitudinal Cell Painting morphology. To demonstrate the framework's utility in a highly challenging real-world scenario characterized by weak signals and sparse reference data, we apply it to a 9-week RPE-1 time course spanning five dose rates (0.003–6.0 mGy/hr). We compute week-matched treated–control morphology deltas and construct evidence-rich prompt payloads that bind (i) quantitative deltas to (ii) retrieved perturbation neighbors, pathway context, and literature snippets via stable evidence identifiers. An LLM generates structured controlled-vocabulary hypotheses at the week$\times$dose level, which are then integrated hierarchically into dose-time phases and global summaries while preserving evidence provenance.

We operationalize auditing with two quantitative tests: V1 citation validity, which verifies that every cited evidence identifier is present in the prompt payload, and V2 proxy-based morphology compatibility, which measures whether predicted process labels are consistent with the top changed morphology features under an explicit proxy table. In the audited run, V1 detected no invalid evidence references, and V2 showed non-trivial morphology compatibility that increased with perturbation strength. At week$\times$dose resolution, V2 also showed a modest positive association with an LLM-external morphology drift summary computed from the same Cell Painting data distribution. Crucially, this auditable pipeline distilled complex trajectories into falsifiable hypotheses, including a distinct adaptive phenotype associated with metabolic reprogramming and proteostatic stress at lower dose rates (0.003-0.3 mGy/hr). Limitations include proxy-based evaluation granularity and the absence of ground-truth mechanism labels. The source code and datasets for this study are publicly available at \url{https://github.com/cellpainting-llm-auditor}.
\end{abstract}


\maketitle

\section{Introduction}

\begin{figure*}[t]
  \centering
  \includegraphics[width=1\textwidth]{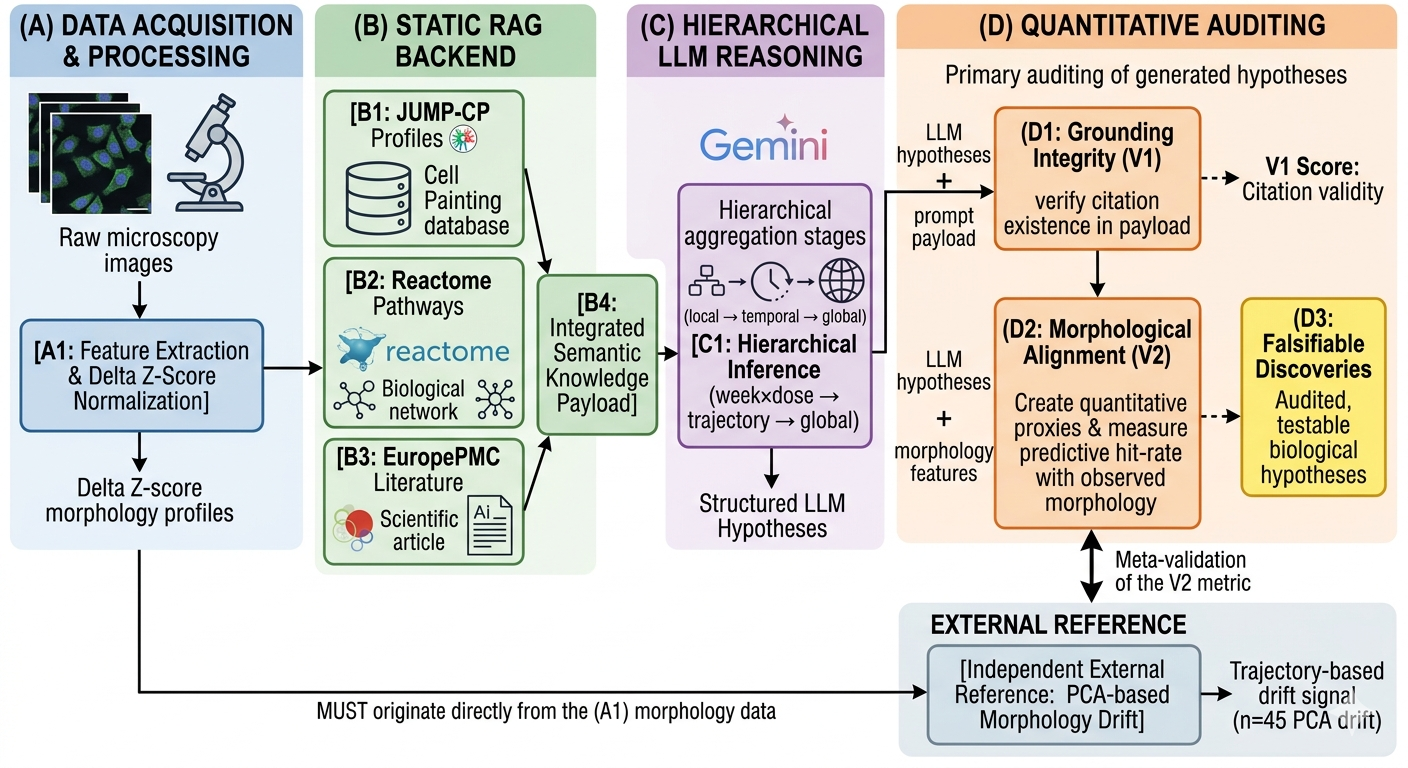}
  \caption{Architectural overview of the evaluation-first LLM-driven morphological reasoning and validation pipeline. The framework consists of four integrated modules: (A) Data Acquisition \& Processing, where raw microscopy images undergo feature extraction and delta Z-score normalization to yield quantitative morphology profiles; (B) Static RAG Backend, which constructs an integrated semantic knowledge payload by synthesizing morphological reference profiles (JUMP-CP), mechanistic biological pathways (Reactome), and scientific literature contexts (EuropePMC); (C) Hierarchical LLM Reasoning, employing the Gemini large language model to iteratively consolidate local condition-specific inferences (week × dose) into temporal trajectories and unified global hypotheses; and (D) Quantitative Auditing, the primary evaluation module where hypotheses are rigorously anchored against grounding integrity (V1: citation validity) and morphological alignment (V2: quantitative predictive hit-rate) to produce audited, falsifiable biological discoveries. Furthermore, the V2 auditing metric is meta-validated against an independent external reference of trajectory-based PCA morphology drift. (Note: The visual representation of this diagram was generated with the assistance of the Gemini.)}
  \label{fig:pipeline}
\end{figure*}
  
High-content imaging assays provide a sensitive readout of how perturbations reshape cellular state. Cell Painting--style morphological profiling converts multiplexed fluorescence images into quantitative single-cell feature vectors spanning size, shape, intensity, and texture~\cite{Bray2016CellPainting}. Repeated measurement enables longitudinal trajectories, but mechanistic interpretation is particularly challenging for weak, chronic perturbations such as low-dose and low-dose-rate ionizing radiation, where responses can be multifactorial and context dependent~\cite{UNSCEAR2020LowDose,BEIRVII2006}.

The interpretability bottleneck is largely downstream of feature extraction. Morphological profiles are high-dimensional and sensitive to technical variation, often requiring robust normalization and aggregation before comparisons are meaningful across plates and timepoints~\cite{caicedo2017data,JUMP2024CellPainting}. Even after preprocessing, translating a trajectory in morphology space into biological hypotheses typically remains manual. Public reference resources partially mitigate this gap via ``profile matching,'' where new signatures are compared to large perturbation libraries (e.g., JUMP) to suggest candidate gene programs or pathways~\cite{JUMP2024CellPainting}. However, similarity alone rarely yields an end-to-end explanation---especially in longitudinal settings where dominant processes may shift by dose or phase.

Large language models (LLMs) offer an appealing interface for synthesizing heterogeneous evidence, but scientific deployment requires auditing: fluent narratives can be unfaithful to evidence and may include fabricated or incorrect citations~\cite{LLMHallucinationCitations,HumanCitationErrors}. Retrieval-augmented generation (RAG) improves provenance by coupling generation to external evidence stores~\cite{Lewis2020RAG}, and curated resources such as Reactome provide standardized pathway context~\cite{Reactome2024}.

In this work, we introduce an LLM-centered, retrieval-augmented interpretation framework for longitudinal Cell Painting morphology that is evaluation-first by design. To demonstrate the practical utility of this framework in a highly challenging real-world scenario, we apply it to the longitudinal profiling of chronic low-dose-rate (LDR) ionizing radiation—a domain where morphological signals are notoriously weak and established reference mechanisms are scarce. Rather than treating LLM narratives as conclusions, we treat them as auditable scientific artifacts: structured hypotheses grounded in (i) measured morphology deltas, and (ii) retrieved external evidence (perturbation neighbors, pathway context, literature snippets).

Specifically, we operationalize auditing through two rigorous quantitative tests. First, we enforce citation integrity (V1) by verifying that every piece of evidence cited by the LLM strictly maps back to the provided prompt payload, effectively eliminating reference hallucinations. Second, we assess biological fidelity via a proxy-based alignment metric (V2), which evaluates the consistency between the LLM's predicted biological processes and the most prominent morphological feature shifts. To further substantiate these findings, we establish an independent, morphology-only PCA drift summary. This drift metric serves as a crucial secondary quantitative anchor, ensuring that the semantic trajectories proposed by the LLM are fundamentally grounded in the principal axes of phenotypic variation observed in the raw data (Figure~\ref{fig:pipeline}).

\vspace{0.5em}
\noindent \textbf{Contributions:}
\begin{itemize}
  \item \textbf{Evaluation-first auditing for LLM interpretation of morphology:} Formal validation of hypothesis grounding (V1) and quantitative proxy alignment (V2), designed specifically for settings lacking ground-truth mechanism labels.
  \item \textbf{Traceable retrieval-augmented prompting:} A robust prompt payload schema that binds quantitative morphology deltas to retrieved perturbation and literature evidence via stable IDs, enabling automatic citation verification.
  \item \textbf{Hierarchical reasoning with provenance:} A multi-step synthesis approach (dose-time and global summaries) that preserves longitudinal structure and reliably propagates evidence references, preventing the loss of context typical of flattened week\,$\times$\,dose outputs.
\end{itemize}

\section{Related Work}
\subsection{Cell Painting and morphological profiling}
Cell Painting is a widely adopted foundation for image-based morphological profiling, yielding multi-channel fluorescence images that can be summarized into quantitative single-cell feature vectors \cite{Bray2016CellPainting}. A recurring theme in the literature is that interpretability is limited less by feature extraction than by downstream processing and contextualization: profiles often require robust normalization, aggregation, and careful similarity computation before they can be compared across batches or experimental conditions \cite{JUMP2024CellPainting}.

Recently, large public reference resources have emphasized both the promise and complexity of morphological profiling. The JUMP Cell Painting Consortium released a benchmark-scale dataset spanning matched chemical and genetic perturbations to support similarity evaluation and representation learning \cite{JUMP2024CellPainting}. While early approaches relied on classical image analysis pipelines, modern efforts increasingly leverage deep representation learning—such as weakly supervised learning and vision transformers—to extract more robust phenotypic embeddings \cite{caicedo2017data,JUMP2024CellPainting}. Such resources enable ``profile matching'' as a practical mechanism-discovery heuristic, but nearest-neighbor similarity typically requires additional pathway and literature context. This remains especially challenging for longitudinal settings where phenotypes evolve dynamically over time.

\subsection{LLMs, RAG systems, and biological interpretation}
Biomedical LLMs have become central tools for text mining and synthesis, building on domain-adapted pretraining (e.g., BioBERT, PubMedBERT) and subsequent generative models tailored for biomedical knowledge tasks \cite{BioBERT,PubMedBERT,BioGPT}. More recently, state-of-the-art models like Med-PaLM have demonstrated expert-level performance on medical and biological reasoning benchmarks, signaling a shift toward highly capable, domain-specific reasoning engines \cite{Singhal2023MedPaLM}. 

In parallel, retrieval-augmented generation (RAG) couples a generator with an evidence store to improve provenance and factuality for knowledge-intensive generation \cite{Lewis2020RAG}. In biology, curated resources such as Reactome provide standardized pathway definitions and gene--pathway mappings that are well suited for evidence-aware interpretation pipelines \cite{Reactome2024}. The integration of RAG into scientific workflows is now moving beyond simple question-answering toward autonomous hypothesis generation and complex data interpretation \cite{Boiko2023Emergent}.

\subsection{Grounding, evaluation, and interpretation auditing}
Despite their capabilities, neural generation in scientific settings can be fluent yet unfaithful to underlying evidence, motivating explicit grounding constraints and automated checks \cite{maynez2020faithfulness,bender2021dangers}. Prior work proposes faithfulness evaluations and truthfulness benchmarks (e.g., TruthfulQA) \cite{lin2022truthfulqa} and scientific claim verification benchmarks (e.g., SciFact) \cite{SciFact}. More broadly, documentation and auditing frameworks (e.g., model cards, datasheets) motivate treating AI artifacts as objects requiring structured reporting and evaluation \cite{ModelCards,Datasheets}.

To the best of our knowledge, no existing computational framework natively integrates longitudinal Cell Painting trajectories with RAG-driven LLM synthesis and quantitative auditing. Traditional morphological profiling largely relies on single-time-point proximity matching (e.g., using JUMP-CP benchmarks), which limits its ability to capture weak, continuous perturbations unfolding over multiple weeks. Conversely, standard RAG applications in biology primarily operate on textual data and lack direct grounding in quantitative, image-derived morphological changes.

Our work bridges this gap by introducing an \emph{evaluation-first auditing} framework for LLM-generated hypotheses in longitudinal morphology. Rather than assuming that retrieval augmentation alone guarantees correctness, we formalize measurable constraints—such as grounding integrity—and introduce quantitative alignment tests, including proxy-based morphology consistency. Crucially, these LLM-centric evaluation metrics are anchored to an independent morphology-only reference used strictly for validation, enabling a new paradigm of auditable morphological reasoning where direct empirical comparison to existing approaches remains challenging.

\section{Methods}
\subsection{Experimental design and dataset}
We studied long-term cellular responses to chronic low-dose ionizing radiation in RPE-1 (human retinal pigment epithelial) cells using the Cell Painting assay \cite{Bray2016CellPainting}. Cells were exposed at five dose rates (mGy/hr): 0.003, 0.03, 0.3, 3.0, and 6.0, and profiled longitudinally from week~1 through week~9 (9 weeks total; Table~\ref{tab:app_dataset}).

\textbf{Paired treated--control design.} For each week, the treated plate (designated as Plate 3) was paired with a same-week control plate (Plate C). Within each week $w$ and dose $d$, we analyzed four treated replicate wells (180 treated wells total). Control wells were defined on Plate~C (2 wells in week~1; 4 wells in weeks~2--9; 34 control wells total), yielding 214 wells analyzed overall (Table~\ref{tab:app_dataset}).

\textbf{Cell-level file structure.} Cell segmentation and feature extraction produced per-image tables of single-cell features. For each well, images were acquired in $f\in\{1,\dots,9\}$ fields of view (FOVs) and $p\in\{1,\dots,15\}$ focal planes (135 images per well nominal). Across the study, 28,886 cell-level tables were analyzed (expected 28,890; four files missing), comprising 6,375,699 single-cell observations (Table~\ref{tab:app_dataset}). Single-cell segmentation was performed using the CellSAM model \cite{marks2025cellsam} to extract instance-level masks prior to morphological feature computation.

\subsection{Morphology preprocessing pipeline (Steps~1--3)}
The preprocessing pipeline converts cell-level observations into well-level treated--control deltas suitable for retrieval and LLM prompting.

\subsubsection{Metadata construction (Step 1).}
We parse the directory structure and file naming convention to map each cell-level file to a well identifier and experimental context $(w,d,\mathrm{condition})$ under the paired Plate~3/Plate~C design. The result is stored as a metadata index table.

\subsubsection{Cell-level to well-level aggregation (Step 2).}
\begin{sloppypar}
For each well, we pool all single-cell observations across its available FOV$\times$plane files and compute robust summary statistics for each base morphology feature: median, median absolute deviation (MAD), 10th percentile, and 90th percentile. We exclude geometry/QC-like columns \texttt{orientation}, \texttt{centroid\_x}, and \texttt{centroid\_y} during aggregation. For the nine base features (\texttt{area}, \texttt{perimeter}, \texttt{eccentricity}, \texttt{solidity}, \texttt{mean\_intensity}, \texttt{glcm\_contrast}, \texttt{glcm\_correlation}, \texttt{glcm\_energy}, \texttt{glcm\_homogeneity}), this yields 36 well-level summary features.
\end{sloppypar}

\subsubsection{Robust within-week normalization and treated--control deltas (Step 3).}
Let $x_{i,f}$ denote the aggregated value for well $i$ and feature $f$, and let $w(i)$ denote the week of well $i$. We perform robust within-week normalization using the per-week feature median and MAD computed across \emph{all} wells (treated and control) in the same week. Letting $W(i)$ denote the set of all wells in week $w(i)$, and with $\epsilon = 10^{-6}$, the normalized value is
\begin{equation}
z_{i,f} = \frac{x_{i,f} - \mathrm{median}_{j \in W(i)}\, x_{j,f}}{\mathrm{MAD}_{j \in W(i)}\, x_{j,f} + \epsilon}.
\label{eq:robust_norm}
\end{equation}
We then define the week-specific control baseline vector as the feature-wise median of normalized control wells $C_w$ in week $w$. The treated--control delta for well $i$ is
\begin{equation}
\Delta_{i,f} = z_{i,f} - \mathrm{median}_{j\in C_{w(i)}}\, z_{j,f}.
\label{eq:delta}
\end{equation}
To ensure stable baselines, we require at least two control wells per week and record control coverage diagnostics. Dataset scale and preprocessing exclusions are summarized in Table~\ref{tab:app_dataset}.

\subsection{LLM-external morphology drift summary (Step~4)}
We compute an LLM-external summary of morphology drift over time for validation-oriented sanity checks only. This module is independent of the retrieval and generation stack, but it is still derived from the same Cell Painting data distribution and is therefore not treated as independent biological validation.

\subsubsection{PCA fitting and drift definition.}
We fit PCA on the normalized well profiles ($z$) restricted to control wells (Plate~C) across all weeks. We use $n_{\mathrm{PC}}=12$ components and apply whitening. We then project each delta profile $\Delta_i$ into this PCA space and define drift magnitude as the Euclidean norm of the whitened PCA score vector:
\begin{equation}
D_i = \left\lVert \mathrm{PCA}_{\mathrm{whiten}}(\Delta_i) \right\rVert_2.
\label{eq:drift}
\end{equation}
For each $(w,d)$, we aggregate drift by averaging $D_i$ across the four treated replicate wells.

\subsubsection{Week$\times$dose and dose-level summaries.}
We summarize drift at two resolutions. First, for each $(w,d)$ we retain the week$\times$dose mean drift as a higher-resolution morphology-only summary. Second, for each dose $d$, we summarize the week-by-week drift trajectory using (i) a linear slope computed by least-squares regression of mean drift against week index, and (ii) an area-under-the-curve (AUC) computed by trapezoidal integration over weeks. These summaries are used only as LLM-external quantitative anchors for sanity checks on score behavior.

\subsection{Retrieval-augmented prompt construction (Steps~5--6)}
We construct a structured evidence payload for each $(w,d)$ to support grounded, traceable LLM hypothesis generation.

\subsubsection{Reduced morphology neighbor index.}
We build a reduced-feature index over a public Cell Painting perturbation reference (JUMP). We compute a 9-dimensional embedding corresponding to the base morphology feature set above by mapping each feature to matching Cell Painting columns and averaging across available channels/scales. Reference embeddings are z-scored using mean and standard deviation with $\epsilon=10^{-8}$. The index uses exact L2 neighbor search (FAISS when available), otherwise a scikit-learn nearest-neighbor backend with Euclidean distance \cite{johnson2019billion,pedregosa2011scikit}.

\subsubsection{Neighbor search and selection.}
For each $(w,d)$ observation, we query the reference index with $k=100$ candidate neighbors and retain up to 10 for prompting, enforcing that at least three neighbors correspond to gene perturbations when possible. Neighbor records are assigned stable evidence IDs of the form \texttt{jump:<jump\allowbreak\_doc\_id>}.

\subsubsection{Reactome pathway retrieval.}
We construct a retrieval index over Reactome pathway descriptions using TF--IDF (1--2 grams; cosine similarity; \texttt{min\_df}=2; \texttt{max\_features}=\allowbreak 250000) \cite{Reactome2024}. For each $(w,d)$ payload, we aggregate enriched pathways for Entrez genes associated with the candidate neighbors and retain the top 12 pathways by frequency. For dose-time and global aggregation (Step~8), pathway references can be cited using \texttt{path:<pathway\allowbreak\_id\_or\_name>} and are validated against the payload; for week$\times$dose generation (Step~7), enforceable retrieval IDs are limited to \texttt{jump:} and \texttt{lit:} (see below).

\subsubsection{Literature snippets and quantitative observations.}
When enabled, we retrieve brief literature snippets per prompt via the Europe PMC REST API \cite{europe2015europe}, using radiation-focused queries augmented with the cell line and additional terms (\texttt{dose rate}, \texttt{protracted}, \texttt{long-term}). We retrieve up to 10 hits per query and optionally pull and truncate selected full-text sections when available (sections \texttt{methods}, \texttt{results}, \texttt{discussion}, \texttt{supplementary}; section character cap 5000). Each snippet is assigned a stable ID \texttt{lit:q<query\_idx>:\allowbreak<lit\_id>}. Each payload also includes a quantitative observation block with a unique ID \texttt{obs\_w<week>\allowbreak\_d<dose>} and the top changed morphology features by absolute mean delta.

\subsubsection{Prompt payload schema and traceability.}
Each JSONL prompt record stores: (i) the observation block (obs ID and top-changed features), (ii) selected neighbors (with \texttt{jump:} IDs), (iii) literature snippets (with \texttt{lit:} IDs), and (iv) pathway evidence (pathway IDs). These evidence identifiers form the reference set used for grounding validation (V1).

\subsection{Structured LLM hypothesis generation (Step~7)}
We run an LLM reasoner per $(w,d)$ prompt with strict structured-output constraints.

\subsubsection{Model and decoding.}
We use Google's Gemini model via Vertex AI \cite{team2023gemini} with temperature 0.2, thinking level \texttt{HIGH}, maximum output length 8192 tokens, and a prompt length cap of 300,000 characters. The runner retries up to three times on schema or grounding failures. The specific model revision served by the endpoint was \texttt{gemini-3-pro-preview}.

\subsubsection{JSON-only schema and controlled vocabulary.}
\begin{sloppypar}
The model is instructed to output JSON only, conforming to a schema with one or more hypotheses per prompt. Each hypothesis includes: \texttt{hypothesis\_id}, \texttt{biological\_process}, \texttt{summary}, \texttt{quant\_evidence\_refs}, \texttt{retrieval\_refs}, \texttt{testable\_predictions}, \texttt{confidence}, and a categorical \texttt{process\_label} constrained to an 11-label controlled vocabulary (Table~\ref{tab:app_system}). Schema validation is enforced programmatically.
\end{sloppypar}

\subsubsection{Grounding enforcement.}
\begin{sloppypar}
For week$\times$dose hypotheses, grounding validation requires that all \texttt{quant\_evidence\_refs} are of the form \texttt{obs:<obs\_id>} and match the prompt's observation ID, and that all \texttt{retrieval\_refs} are of the form \texttt{jump:<jump\_id>} or \texttt{lit:<lit\_id>} and appear in the payload. We also require that each hypothesis cites at least one observation and one retrieved item.
\end{sloppypar}

\subsection{Hierarchical integration across scales (Step~8)}
Longitudinal interpretation benefits from hierarchical reasoning that preserves temporal structure.

\subsubsection{Dose-time aggregation.}
We construct dose-time prompts by grouping week$\times$dose outputs across weeks for each dose and adding a timepoint catalog that retains per-week observation IDs, selected neighbors, literature snippets, and pathway evidence. A second LLM pass produces phase-level summaries (e.g., early/mid/late) with evidence references that must cite at least one observation and one retrieved item per phase.

\subsubsection{Why week $\times$ dose first (and why not flat aggregation).}
\begin{sloppypar}
Week$\times$dose reasoning is performed first because it is the smallest unit in which (i) quantitative deltas $\Delta_{i,f}$ are directly defined against the matched weekly control baseline, and (ii) retrieval evidence (nearest neighbors, pathway context, literature snippets) is conditioned on the same quantitative signature. Aggregating all timepoints into a single prompt would mix heterogeneous temporal regimes and can obscure non-monotonic dynamics: feature directions can cancel when averaged across weeks, phase-specific patterns can be lost, and trends can reverse when marginalizing over time (a Simpson's paradox). Flat prompts also dilute provenance by coupling a large evidence pool to a small number of claims, increasing ambiguity about which observations support which statements.
\end{sloppypar}

To preserve traceability, we keep evidence identifiers stable across scales. Each week$\times$dose prompt assigns a unique observation ID (e.g., \texttt{obs:obs\_w\{week\}\allowbreak\_d\{dose\}}) and retains retrieval IDs (\texttt{jump:}, \texttt{lit:}, and Reactome IDs). Dose\_time prompts embed a per-week catalog of these IDs; global prompts embed the per-dose catalogs. Downstream, V1 grounding checks operate on these identifier sets to verify that every phase- or global-level statement cites concrete week-level measurements and retrieved evidence.

\subsubsection{Global summarization.}
We further summarize across doses by constructing a global prompt that preserves the dose-time evidence catalog and produces a small set of global hypotheses with evidence references. Hierarchical reasoning avoids flattening heterogeneous timepoints and enables phase-specific claims while maintaining provenance.

\subsection{Validation framework}
We distinguish (i) measured morphology quantities ($\Delta$, drift), (ii) retrieval evidence (neighbors, snippets, pathways), and (iii) LLM-generated hypotheses. We quantify two complementary validation layers.

\subsubsection{V1: citation validity (grounding integrity).}
V1 checks citation integrity: do evidence references in an LLM hypothesis correspond to evidence IDs present in the associated prompt payload? Let $\mathcal{O}(p)$ be the observation ID set for payload $p$ and $\mathcal{E}(p)$ the set of retrieval IDs (\texttt{jump:} and \texttt{lit:} for week$\times$dose). For a hypothesis $h$ with reference sets $\mathcal{Q}(h)$ (quantitative) and $\mathcal{R}(h)$ (retrieval), we define
\begin{equation}
\begin{aligned}
\mathrm{V1}(h,p) = \mathbb{I}\Big[ &\mathcal{Q}(h)\subseteq \mathcal{O}(p)\ \wedge\ \mathcal{R}(h)\subseteq \mathcal{E}(p)\ \wedge \\
&|\mathcal{Q}(h)|>0\ \wedge\ |\mathcal{R}(h)|>0 \Big].
\end{aligned}
\label{eq:v1}
\end{equation}
V1 therefore evaluates identifier-level provenance validity within a closed evidence set. It does not test whether a cited item semantically entails or fully supports a claim.

\subsubsection{V2: proxy-based process--morphology compatibility.}
\begin{sloppypar}
V2 tests whether a predicted \texttt{process\_label} is compatible with the prompt's quantitative morphology signature. To operationalize this comparison, we use an explicit proxy mapping between the controlled vocabulary of biological processes and expected directional shifts in core morphological features (Table~\ref{tab:s5_proxy_map}). Because morphological profiles are highly multiplexed, we restricted our proxy definitions to well-established, macroscopic cellular hallmarks. The mapping is intentionally simplified, but it is fully specified in the manuscript and programmatically applied, making the evaluation transparent and reproducible.
\end{sloppypar}

For instance, cellular senescence is classically characterized by cellular hypertrophy (significantly increased area and perimeter) alongside increased cytoplasmic granularity and vacuolization (reflected as increased \texttt{glcm\_contrast} and decreased \texttt{glcm\_energy})~\cite{HernandezSegura2018Senescence}. Conversely, apoptosis and cell death typically manifest as severe cell shrinkage (decreased area) and chromatin condensation or membrane blebbing, which drive high textural contrast~\cite{Kroemer2009Apoptosis}. Furthermore, cell-cycle arrest frequently leads to continued macromolecular synthesis without division, resulting in increased overall cell size~\cite{Neurohr2019CellSize}. Processes such as oxidative stress and DNA damage response, while lacking a uniform macroscopic size shift, frequently induce cytoskeletal remodeling and nuclear texture alterations identifiable via GLCM metrics~\cite{Rohban2017Morphological}. 

We emphasize that this predefined proxy mapping is not intended as a definitive single-cell mechanism classifier or a ground-truth benchmark. Rather, it serves as a conservative, literature-grounded consistency check designed to penalize biologically contradictory LLM outputs (e.g., an LLM predicting ``apoptosis'' when the observation payload clearly indicates massive cell enlargement).

\begin{sloppypar}
Formally, for each label $\ell$, let $\mathcal{P}_{\ell}$ be its defined proxy set over the reduced feature vocabulary. Let $\mathcal{T}$ denote the set of top features provided in the prompt (ranked by $|\Delta|$). We define Hit@10 as the fraction of top features belonging to the proxy set:
\end{sloppypar}
\begin{equation}
\mathrm{Hit@10}(\ell) = \frac{|\mathcal{T} \cap \mathcal{P}_{\ell}|}{|\mathcal{T}|}.
\label{eq:hitk}
\end{equation}
We also compute a magnitude-aware weighted proxy score as the fraction of total absolute delta mass in $\mathcal{T}$ explained by proxy hits:
\begin{equation}
\mathrm{WP}(\ell) = \frac{\sum\limits_{f\in \mathcal{T}\cap \mathcal{P}_{\ell}} |\Delta_f|}{\sum\limits_{f\in \mathcal{T}} |\Delta_f|}.
\label{eq:weighted_proxy}
\end{equation}

\subsubsection{Consistency with the drift summary.}
To relate LLM-centric validation to an LLM-external morphology-only signal, we perform two correlation analyses. First, at week$\times$dose resolution we aggregate per-hypothesis V2 scores to the observation level by mean (with median and max summaries reported in supplementary diagnostics) and correlate these summaries with the corresponding week$\times$dose drift mean from Step~4. Second, we compute a coarser dose-level comparison using per-dose V2 summaries and dose-level drift summaries (mean drift, slope, and AUC). Pearson and Spearman correlations are reported with nonparametric bootstrap uncertainty (10,000 resamples). These analyses are intended as sanity checks on score behavior rather than as independent biological validation.

\subsubsection{Supplementary diagnostics.}
Additional quantitative and retrieval diagnostics referenced throughout the paper are provided in Supplementary Materials in Appendix.

\section{Results}
\subsection{Grounding integrity (V1)}
Across the full longitudinal design (9 weeks $\times$ 5 dose rates), the week$\times$dose LLM reasoning stage produced 45/45 successful prompt records (0 failures), yielding 136 total hypotheses. Under the implemented citation-integrity check, no grounding failures were detected: 0 invalid evidence-reference cases and 0 records missing grounding statistics.

V1 grounding validity was perfect under strict grounding criteria (Methods, Eq.~\ref{eq:v1}). Both the mean observation-reference validity rate and the mean retrieval-reference validity rate were 1.0. Furthermore, all hypotheses contained at least one observation reference and one retrieval reference (coverage rate = 1.0). Aggregate V1 metrics are summarized in Table~\ref{tab:validation}; additional reference-count diagnostics are provided in Table~\ref{tab:s4_grounding} and Figure~\ref{fig:s2_ref_comp}. By construction, these V1 results establish provenance validity within the closed prompt evidence set; they do not by themselves show that a cited item semantically supports a claim.

\subsection{Representative auditable hypotheses}
To make the audit-first outputs concrete, we highlight one representative week$\times$dose hypothesis alongside its supporting quantitative signature. Table~\ref{tab:rep_payload} displays the input prompt payload for week~4 at 3.0~mGy/hr, which reports the top-$|\Delta|$ morphology deltas (group-mean $\Delta Z$; values rounded to three decimals).

\begin{table}[ht!]
  \caption{Quantitative prompt payload (top morphology deltas) for week 4 at 3.0 mGy/hr.}
  \label{tab:rep_payload}
  \centering
  \small
  \begin{tabular}{@{} l r @{}}
  \toprule
  \textbf{Feature} & $\boldsymbol{\Delta Z}$ \\
  \midrule
  \texttt{mean\_intensity\_median} & $+2.145$\\
  \texttt{glcm\_energy\_median} & $-1.786$\\
  \texttt{solidity\_median} & $+1.364$\\
  \texttt{glcm\_homogeneity\_median} & $-1.204$\\
  \texttt{area\_median} & $+1.007$\\
  \texttt{glcm\_correlation\_median} & $-0.756$\\
  \texttt{eccentricity\_median} & $+0.621$\\
  \texttt{glcm\_contrast\_median} & $-0.268$\\
  \bottomrule
  \end{tabular}
\end{table}

\begin{sloppypar}
Based on this signature and the retrieved context, the LLM generated hypothesis \texttt{hyp\_w4\_d3.0\_1}, classifying the state as \texttt{senescence\_like} with high confidence (Table~\ref{tab:rep_hypothesis}). We present this example as an auditable biological hypothesis: its cited evidence is traceable, and its declared label can be checked for morphology-compatibility under V2.
\end{sloppypar}

\begin{table}[ht!]
  \caption{Representative week$\times$dose hypothesis generated by the LLM.}
  \label{tab:rep_hypothesis}
  \centering
  \small
  \begin{tabular}{@{} l p{0.70\columnwidth} @{}}
  \toprule
  \textbf{Field} & \textbf{Content} \\
  \midrule
  \textbf{Hypothesis ID}  & \texttt{hyp\_w4\_d3.0\_1} \\
  \textbf{Process Label}  & \texttt{senescence\_like} (Confidence: \texttt{high}) \\
  \textbf{Summary}        & ``The observed increase in cell area and texture heterogeneity, characteristic of cellular hypertrophy, aligns with a senescence-like state. This is supported by morphological similarity to the perturbation of RRAGD, a regulator of mTOR signaling, which is a key driver of senescent cell enlargement and metabolic activity.'' \\
  \midrule
  \textbf{Quant Evidence} & \texttt{obs:obs\_w4\_d3.0} \\
  \textbf{Retrieval Refs} & \texttt{jump:row\_916481}, \texttt{lit:q1:4}, \texttt{lit:q1:6} \\
  \bottomrule
  \end{tabular}
\end{table}

\begin{sloppypar}
This example demonstrates identifier-level citation validity and strong proxy compatibility under our evaluation framework. It passes V1 because all cited evidence identifiers are strictly present in the associated payload (record-level validity: \texttt{obs\_valid\_rate=1.0}, \texttt{retrieval\_valid\_rate=1.0}; 0 invalid references). Furthermore, it passes V2 proxy compatibility: under the \texttt{senescence\_like} mapping (Table~\ref{tab:s5_proxy_map}), 7 top features are proxy hits ($\mathrm{Hit@10}=0.875$), and these hits account for 86.8\% of the absolute delta mass ($\mathrm{WP}=0.868$; Methods, Eq.~\ref{eq:hitk}--\ref{eq:weighted_proxy}). This does not by itself prove that the cited evidence semantically entails the claim, but it shows that the hypothesis is auditable under both V1 and V2. A complete structured example of a global-level claim is provided in Table ~\ref{tab:global_claim_example}.
\end{sloppypar}

\subsection{Proxy alignment performance (V2)}
We next evaluated whether the LLM-generated \texttt{process\_label} hypotheses were quantitatively compatible with observed morphology deltas using the predefined proxy mapping (Methods, Eq.~\ref{eq:hitk}--\ref{eq:weighted_proxy}; proxy definitions in Table~\ref{tab:s5_proxy_map}). At the global (dose$\times$label) level, the mean Hit@10 over rows was 0.4335 ($n_{\mathrm{dose\_label\_rows}}=14$), with 0 missing-delta records.

Dose-stratified results showed a clear dose dependence in best-case compatibility. The per-dose best mean Hit@10 values were 0.2361 (0.003~mGy/hr), 0.4306 (0.03~mGy/hr), 0.4861 (0.3~mGy/hr), 0.8750 (3.0~mGy/hr), and 0.8889 (6.0~mGy/hr).

\begin{figure}[ht!]
  \centering
  \includegraphics[width=\columnwidth]{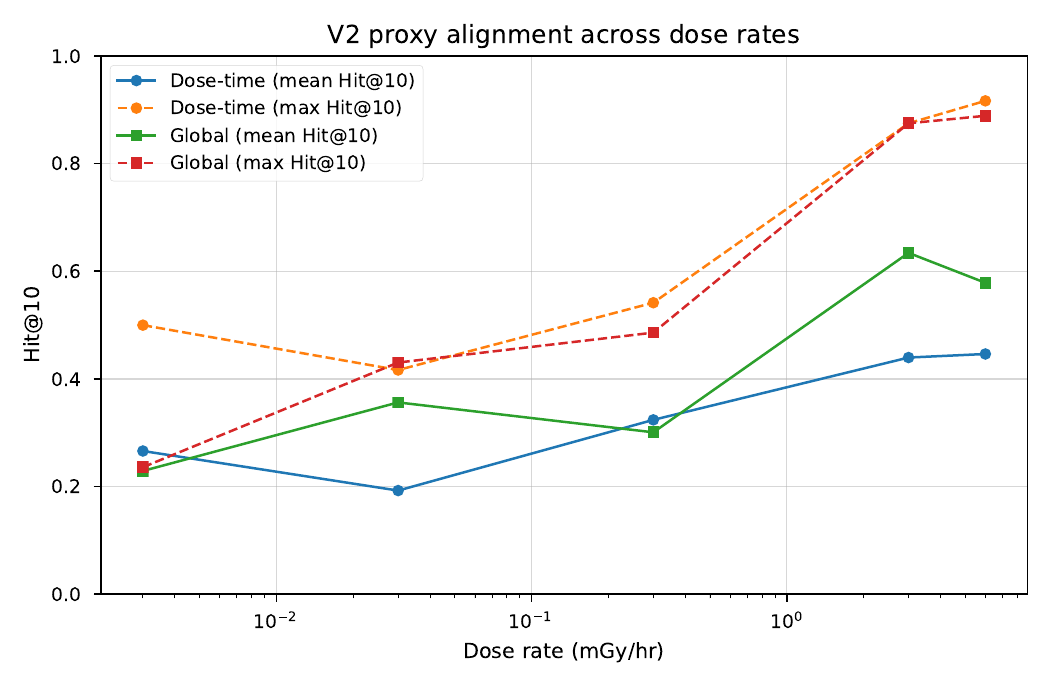}
  \caption{V2 proxy alignment dose-response (Hit@10).}
  \Description{Line plot of V2 Hit@10 versus dose rate for dose\_time and global summaries, showing mean and max.}
  \label{fig:v2_dose_response}
\end{figure}

\begin{figure*}[ht!]
  \centering
  \includegraphics[width=\textwidth]{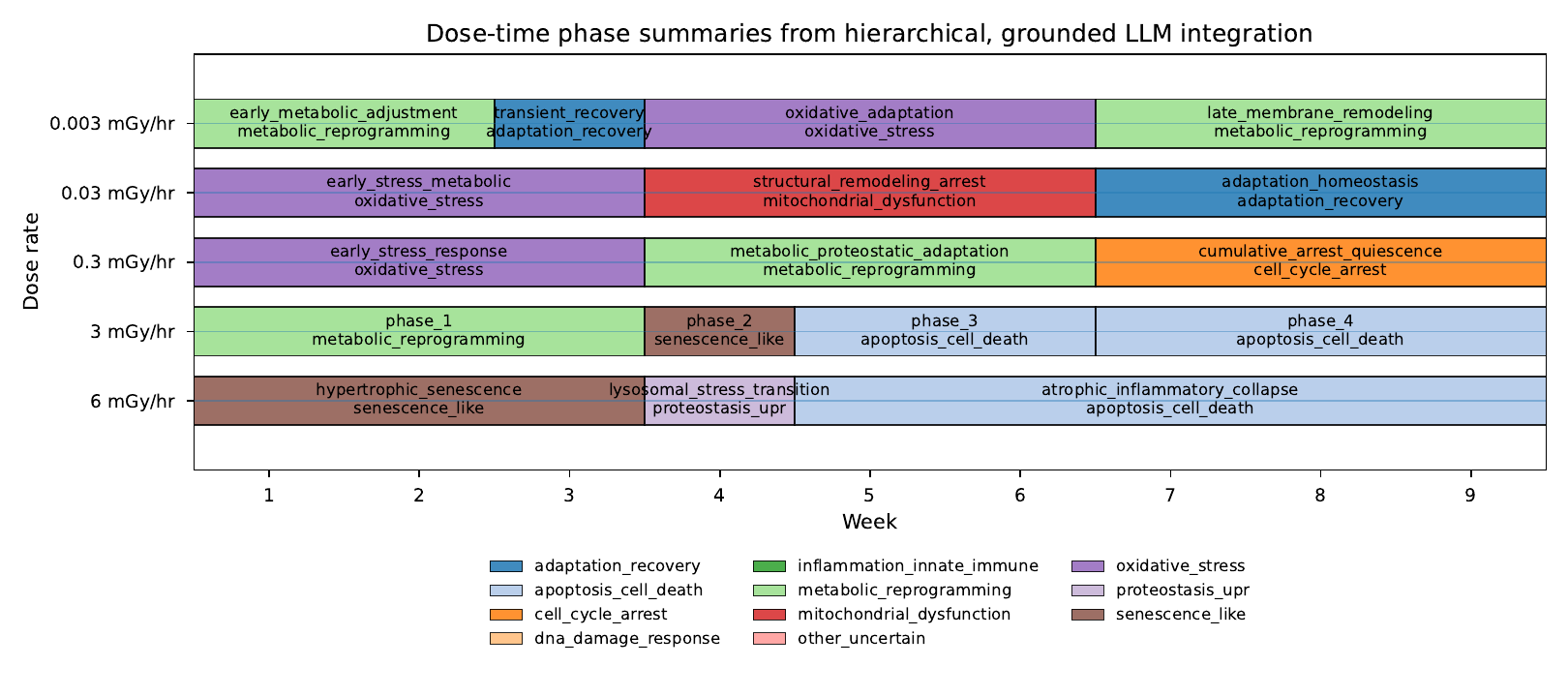}
  \caption{Dose-time integrated phase timeline (hierarchical reasoning output).}
  \Description{Timeline plot showing weeks grouped into phases for each dose, annotated with dominant process labels.}
  \label{fig:dose_time_timeline}
\end{figure*}

Figure~\ref{fig:v2_dose_response} visualizes how V2 alignment varies with dose rate for both hierarchical scales (dose-time and global), reporting both the mean Hit@10 across rows and the best-case (max) Hit@10 at each dose. Quantitatively, the global best-case Hit@10 increases from 0.2361 at 0.003~mGy/hr to 0.8889 at 6.0~mGy/hr, while the global mean Hit@10 increases from 0.2292 to 0.5787 over the same range. Dose\_time summaries show a similar upward trend (mean Hit@10 from 0.2662 to 0.4464; max Hit@10 from 0.5000 to 0.9167). In the evaluation-first framing, this dose-response matters because it indicates that the compatibility metric is sensitive to perturbation strength: when quantitative deltas are weak, proxy alignment is limited, whereas stronger perturbations permit higher compatibility between declared \texttt{process\_label}s and observed morphology signatures.

\begin{table}[ht!]
  \caption{Validation summary (V1 grounding, V2 proxy alignment).}
  \label{tab:validation}
  \centering
  \small
  \begin{tabular}{@{} l r l @{}}
\toprule
\textbf{Metric} & \textbf{Value} & \textbf{Notes} \\
\midrule
\multicolumn{3}{@{}l}{\textbf{V1 grounding integrity}}\\
\addlinespace
Week$\times$dose records (successful) & 45 & Total prompts processed \\
Hypotheses generated & 136 & Across all records \\
Invalid evidence-reference cases & 0 & V1 citation failures \\
Mean obs-reference validity & 1.000 & Valid \texttt{obs:<id>} / total \\
Mean retrieval-reference validity & 1.000 & Valid \texttt{jump:}/\texttt{lit:} / total \\
Coverage (both obs + retrieval refs) & 1.000 & Fraction of hypotheses \\
\midrule
\multicolumn{3}{@{}l}{\textbf{V2 proxy alignment}}\\
\addlinespace
Global mean Hit@10 (over rows) & 0.4335 & Dose$\times$label rows ($n=14$) \\
Missing-delta records & 0 & Should be 0 \\
\addlinespace
\multicolumn{3}{@{}l}{\textit{Per-dose best mean Hit@10 (best row per dose)}}\\
\quad Dose 0.003 mGy/hr & 0.2361 & \\
\quad Dose 0.03 mGy/hr & 0.4306 & \\
\quad Dose 0.3 mGy/hr & 0.4861 & \\
\quad Dose 3.0 mGy/hr & 0.8750 & \\
\quad Dose 6.0 mGy/hr & 0.8889 & \\
\bottomrule
\end{tabular}
\end{table}

\subsection{Hierarchical longitudinal summaries}
Beyond week$\times$dose hypotheses, the system produces hierarchical summaries at the dose\_time and global levels while preserving evidence IDs for traceability (Methods). Dose-time summaries partition each dose trajectory into a small set of temporal phases with dominant process labels and supporting evidence.

Figure~\ref{fig:dose_time_timeline} shows the dose-time integration output as a phase timeline for each dose rate, where contiguous week ranges are grouped into 3--4 phases and assigned dominant \texttt{process\_labels}. This visualization is not used as a biological claim; rather, it demonstrates the intended behavior of hierarchical reasoning: the model produces compact, temporally structured summaries that preserve within-dose phase structure instead of collapsing all weeks into a single undifferentiated narrative. Because phases cite week-level observation IDs and retrieval IDs, they remain auditable under V1 and quantifiable under V2.

\begin{figure}[ht!]
  \centering
  \includegraphics[width=\columnwidth]{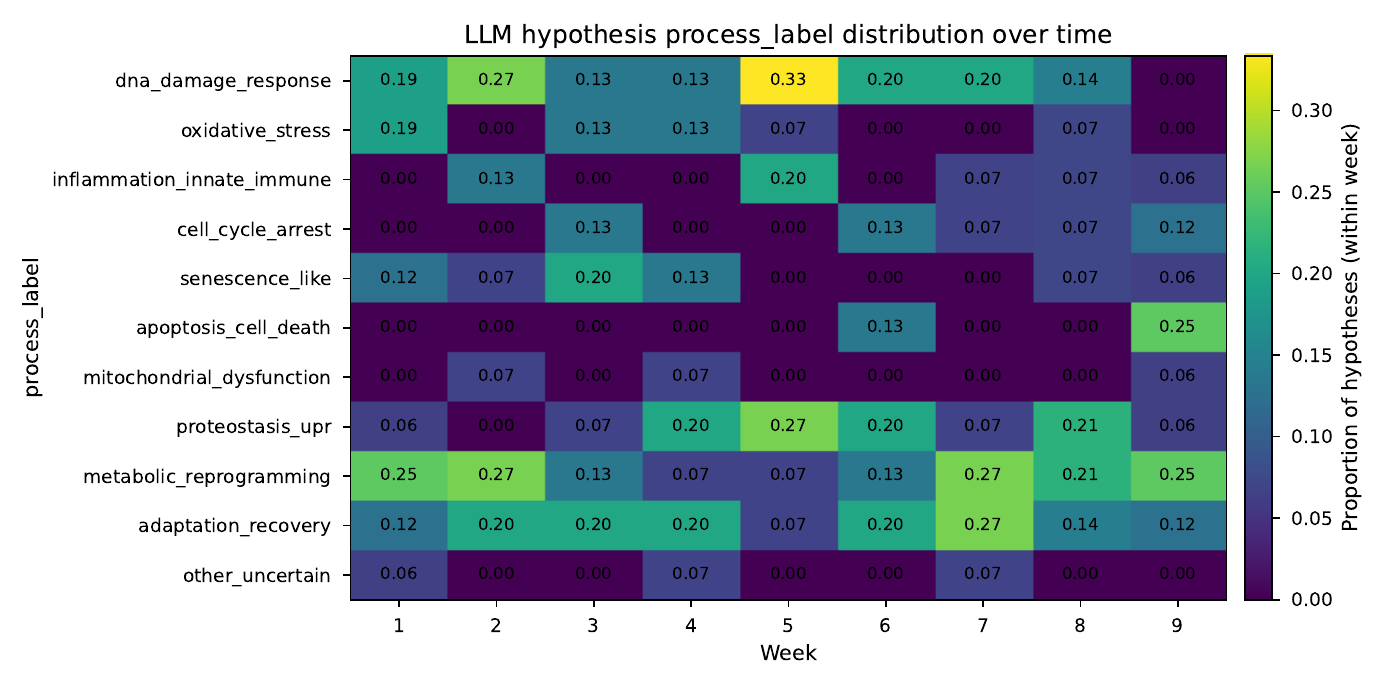}
  \caption{Temporal heatmap of LLM \texttt{process\_label} assignments.}
  \Description{Heatmap of process label counts over weeks and dose rates, showing temporal variation in label usage.}
  \label{fig:process_label_heatmap}
\end{figure}

\begin{sloppypar}
Figure~\ref{fig:process_label_heatmap} summarizes the within-week distribution of \texttt{process\_label} assignments across all week$\times$dose hypotheses. The non-uniform label usage over time provides a descriptive check that the controlled vocabulary is used in a temporally structured manner (rather than collapsing to a single default label). In the evaluation-first framing, this matters because it motivates why phase-aware aggregation (Figure~\ref{fig:dose_time_timeline}) is preferable to flat averaging: the label distribution can shift across weeks, and hierarchical summaries provide a traceable representation of these shifts.
\end{sloppypar}

\subsection{Higher-resolution consistency with an LLM-external morphology drift summary}
To test whether V2 behaves like a non-arbitrary score, we compared it with the Step~4 morphology drift summary computed without any LLM or retrieval components. We first performed a week$\times$dose analysis by aggregating per-hypothesis V2 scores to the corresponding observation level and comparing them with the matched week$\times$dose drift mean ($n=45$ observations). For the mean weighted-proxy summary, the association was positive but modest (Pearson $r=0.307$ with 95\% bootstrap CI $[-0.001, 0.566]$; Spearman $\rho=0.293$ with 95\% CI $[-0.021, 0.558]$; bootstrap $n_{\mathrm{boot}}=10000$, seed 42). Mean Hit@10 showed a similar but slightly weaker pattern (Pearson $r=0.282$; Spearman $\rho=0.265$).

\begin{figure}[ht!]
  \centering
  \includegraphics[width=\columnwidth]{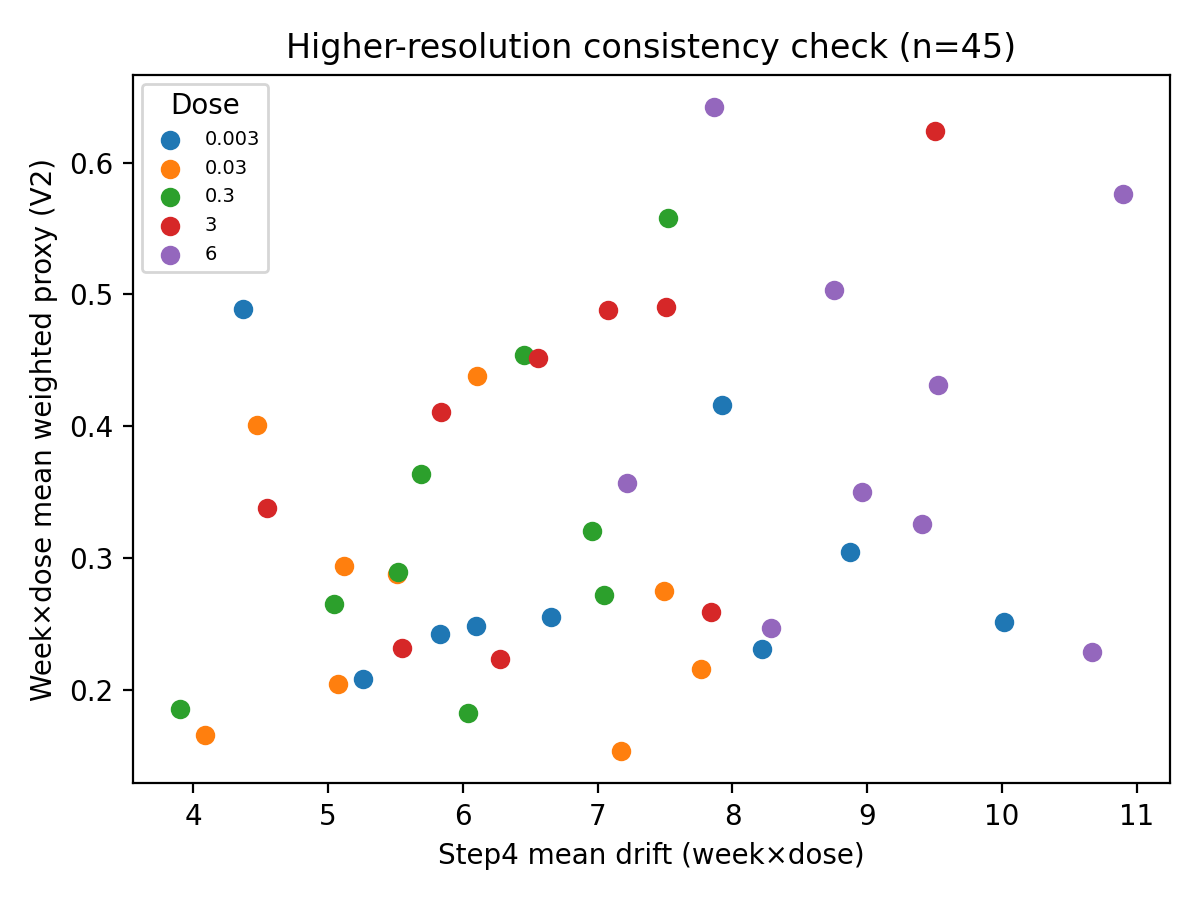}
  \caption{Higher-resolution consistency check between week$\times$dose V2 and the LLM-external morphology drift summary.}
  \Description{Scatter plot of week-by-dose mean weighted proxy score versus week-by-dose morphology drift, with points colored by dose. Each point corresponds to one week-by-dose observation.}
  \label{fig:weekdose_drift_validation}
\end{figure}

Figure~\ref{fig:weekdose_drift_validation} shows this higher-resolution comparison for the mean weighted-proxy summary. The effect size is intentionally interpreted cautiously: both quantities derive from the same Cell Painting distribution, and the purpose of this analysis is only to test whether V2 co-varies with an LLM-external morphology summary rather than behaving as an arbitrary label-matching score (See Appendix \ref{sec:app_dose_sanity} for a complementary coarse dose-level sanity check.).

\subsection{Biological interpretation under quantitative constraints.}

To provide a more concrete biological interpretation, we explicitly examine how LLM-generated hypotheses vary across dose regimes and how these patterns relate to observed morphological signatures.

At higher dose rates (3.0–6.0 mGy/hr), the dominant hypotheses consistently correspond to senescence-like and DNA damage response (DDR)-related processes. This regime is characterized by strong and coherent morphological shifts, including increases in cell area and perimeter along with pronounced changes in texture features. Such changes are consistent with stress-associated morphological remodeling and hypertrophic phenotypes commonly observed during cellular senescence and growth dysregulation~\cite{hernandez2018hallmarks,neurohr2019excessive}. 

Importantly, prior work in image-based profiling has demonstrated that morphological features extracted from Cell Painting assays can reliably capture biologically meaningful variation and map to underlying cellular processes~\cite{caicedo2017data,rohban2017systematic}. In line with this, we observe substantially higher V2 proxy alignment scores in this high-dose regime, indicating strong quantitative consistency between predicted biological processes and observed morphological changes.

In contrast, at lower dose rates (0.003–0.3 mGy/hr), the model produces more heterogeneous hypotheses, including processes such as metabolic reprogramming, proteostatic stress, and adaptive response programs. In this regime, morphological changes are comparatively subtle and less dominated by canonical stress signatures, leading to lower proxy alignment scores and reduced interpretability of process–morphology relationships. 

This pattern is consistent with prior radiobiology literature, which suggests that chronic low-dose radiation often induces non-linear and adaptive cellular responses rather than acute DNA damage-driven phenotypes~\cite{UNSCEAR2020LowDose,BEIRVII2006}.

\section{Discussion}
This work is motivated by a practical gap in high-content biology: morphology is information-rich yet difficult to interpret, while LLMs can generate fluent interpretations that require rigorous auditing before scientific use. We therefore frame LLM outputs as \emph{auditable artifacts}---structured hypotheses with explicit evidence pointers---and emphasize evaluation-first validation via citation validity (V1) and proxy-based morphology compatibility (V2).

\textbf{Scope and limits of quantitative auditing.} V1 enforces \emph{identifier-level} grounding: hypotheses may only cite evidence items that are actually present in the prompt payload. This directly targets out-of-payload references, a known failure mode of neural generation in scientific settings~\cite{lin2022truthfulqa,SciFact,maynez2020faithfulness}. However, V1 does not guarantee that a cited item semantically supports the claim; it is a necessary but not sufficient condition for scientific faithfulness. V2 complements this by evaluating whether predicted \texttt{process\_label}s are compatible with the measured morphology deltas under a predefined proxy mapping. This is intentionally conservative: it does not require that the label is \emph{true} in a mechanistic sense, only that it is \emph{quantitatively compatible} with the observed signature. Because the proxy table is hand-specified, V2 should be interpreted as a transparent and reproducible diagnostic rather than as an objective mechanistic benchmark.

\textbf{Hierarchical reasoning as an auditing strategy.} A core design choice is to generate hypotheses at the most atomic unit of measurement (week\,$\times$\,dose) before aggregation. In longitudinal settings, flat aggregation can obscure phase structure and dilute evidence, and it can yield misleading global summaries when trends differ across subgroups (a form of Simpson's paradox). By preserving evidence IDs from week-level hypotheses into dose\_time phases and global summaries, the system maintains traceability while producing compact narratives suitable for downstream review.

\textbf{Limitations and extensibility.} First, V1 evaluates citation validity but not semantic support; a stronger external assessment would require claim--evidence judgments or expert annotation, which we do not yet have. Second, V2 relies on a predefined proxy mapping from \texttt{process\_label} to expected morphology feature buckets; this mapping is necessarily simplified and may not capture context-dependent morphology--process relationships. Third, the controlled vocabulary constrains expressivity and may collapse distinct biological mechanisms into coarse categories. Fourth, retrieval coverage and bias (e.g., limitations of the reference library, neighbor annotation quality, and pathway text granularity) can shape the evidence available to the LLM and thus the hypotheses produced.

We further emphasize that the use of morphology-only quantitative anchors reflects a practical constraint of the current dataset rather than a limitation of the framework itself. In many longitudinal biological studies---including low-dose radiation experiments---ground-truth mechanism labels and orthogonal modalities are limited or unavailable. Our framework is therefore designed to operate under this constraint, using morphology as a primary quantitative signal while enforcing strict grounding and consistency checks. The auditing structure is inherently extensible: additional modalities such as transcriptomics, proteomics, or targeted biomarkers can be incorporated into the same evidence and validation pipeline to provide stronger mechanistic validation.

Finally, the Step~4 drift summary is derived from the same morphology distribution and should be interpreted only as an LLM-external sanity check, not as independent biological validation.

\textbf{Implications for low-dose radiation studies.} Chronic low-dose and low-dose-rate radiation remains an area where effects can be subtle and context dependent~\cite{UNSCEAR2020LowDose,BEIRVII2006}. Our results do not establish mechanisms; rather, they demonstrate that an LLM-centered interpretation workflow can be instrumented with quantitative auditing so that any proposed biological narrative remains tethered to measured morphology and retrievable evidence.

\textbf{Real-world applicability.}
Our study is conducted on a real-world longitudinal Cell Painting dataset involving chronic low-dose-rate radiation exposure in RPE-1 cells. This setting reflects a particularly challenging regime in radiobiology, where effects are subtle, context-dependent, and lack well-defined ground-truth labels.

In such scenarios, traditional supervised or benchmarking approaches are not feasible. Our framework is specifically designed for this setting: it leverages limited experimental data and external knowledge to generate structured, auditable hypotheses that can guide downstream experimental validation.

This makes the approach directly applicable to other real-world biological contexts characterized by limited labels and complex temporal dynamics, such as drug response profiling, aging studies, and environmental exposure analysis.

\section{Conclusion}
We presented an LLM-centered, retrieval-augmented interpretation framework for longitudinal Cell Painting morphology data that is evaluation-first by design. By enforcing structured JSON outputs with controlled labels and explicit evidence references, and by validating outputs using grounding integrity (V1) and proxy-based process--morphology alignment (V2), the framework enables traceable hypothesis generation that can be quantitatively audited against independent morphology signals.

More broadly, this work contributes a practical template for integrating LLMs into computational biology workflows where narrative interpretation must remain accountable to measured data: hypotheses are generated with explicit provenance, assessed quantitatively, and summarized hierarchically without sacrificing traceability.

\section{Acknowledgment}
This material is based upon works for the LUCID: Low-dose Understanding, Cellular Insights, and Molecular Discoveries program under Award Number DE-AC02-06CH11357 supported by the U.S. Department of Energy, Office of Science, Office of Biological and Environmental Research.

\bibliographystyle{ACM-Reference-Format}
\bibliography{main}

\clearpage
\FloatBarrier

\appendix
\section{Supplementary Materials}
\label{sec:appendix_config}
\FloatBarrier

\setcounter{table}{0}
\renewcommand{\thetable}{A\arabic{table}}

\begin{table}[htbp]
  \centering
  \small
  \caption{Dataset scale and preprocessing summary.}
  \label{tab:app_dataset}
  \begin{tabular}{p{0.32\columnwidth} p{0.26\columnwidth} p{0.36\columnwidth}}
\toprule
\textbf{Item} & \textbf{Value} & \textbf{Notes} \\
\midrule
\multicolumn{3}{l}{\textbf{Biological Setup \& Experimental Design}} \\
\midrule
Cell line & RPE-1 & Human retinal pigment epithelial \\
Assay & Cell Painting & High-content morphology \\
Plate format & 96-well & \\
Time course & Weeks 1--9 & 9 longitudinal time points \\
Dose rates (mGy/hr) & 0.003, 0.03, 0.3, 3.0, 6.0 & 5 treated dose conditions \\
Paired controls & Plate C (weekly) & Paired control plate per week \\
Treated replicates & 4 wells / (week, dose) & Fixed plate map \\
Total treated wells & 180 & 9 weeks $\times$ 5 doses $\times$ 4 wells \\
Total control wells & 34 & Week 1: 2; Weeks 2--9: 4 \\
Total wells analyzed & 214 & Treated + control well profiles \\

\midrule
\multicolumn{3}{l}{\textbf{Data Dimensions}} \\
\midrule
Fields of view (FOV) / well & 9 & $f \in \{1,\ldots,9\}$ \\
Z-planes / FOV & 15 & $p \in \{1,\ldots,15\}$ \\
Nominal CSV files / well & 135 & 9 $\times$ 15 (when complete) \\
Total cell-level CSV files & 28{,}886 & (4 files excluded via QC) \\
Total cell-level observations & 6{,}375{,}699 & Sum of per-well cell counts \\

\midrule
\multicolumn{3}{l}{\textbf{Features \& Preprocessing}} \\
\midrule
Excluded columns & \texttt{orientation}, \texttt{centroid\_x}, \newline \texttt{centroid\_y} & Excluded at well-level aggregation \\
Base morphology feature set & 9 features & \texttt{area}, \texttt{perimeter}, \texttt{eccentricity}, \newline \texttt{solidity}, \texttt{mean\_intensity}, \newline and 4 \texttt{glcm} texture features \\
Well-level summary features & 36 & 9 base $\times$ \{median, MAD, q10, q90\} \\
Within-week normalization & Robust $Z$-score & $(x - \mathrm{median}) / (\mathrm{MAD} + 10^{-6})$ \\
Delta design & Treated -- Control \newline (weekly paired) & $\Delta Z = Z - \mathrm{median}(Z_{\mathrm{control}},\,\mathrm{week})$ \\
\bottomrule
\end{tabular}
\end{table}

\begin{table}[htbp]
  \centering
  \small
  \caption{System configuration summary (retrieval and LLM settings).}
  \label{tab:app_system}
  \begin{tabular}{@{} p{0.22\columnwidth} p{0.30\columnwidth} p{0.43\columnwidth} @{}}
\toprule
\textbf{Component} & \textbf{Parameter} & \textbf{Value} \\
\midrule
\textbf{Retrieval (JUMP)} 
 & Reduced embedding & 9 morphology features; z-score standardization ($\epsilon=10^{-8}$) \\
 & Neighbor search backend & FAISS \texttt{IndexFlatL2} (if available); else sklearn \texttt{NearestNeighbors} (Euclidean) \\
 & Candidate neighbors ($k$) & 100 \\
 & Final neighbors retained & 10 \\
 & Min. gene-like neighbors & 3 \\
\midrule
\textbf{Retrieval (Reactome)} 
 & Gene$\rightarrow$pathway top $N$ & 12 \\
 & Pathway document search & TF--IDF (1--2 grams); cosine similarity \\
 & TF--IDF \texttt{max\_features} & 250,000 \\
 & TF--IDF \texttt{min\_df} & 2 \\
 & Document retrieval $k$ & 6 \\
 & Reactome release & Downloaded from \texttt{/current/} endpoint \\
\midrule
\textbf{LLM Reasoner} 
 & Model & \texttt{gemini-3-pro-preview} \\
 & Temperature & 0.2 \\
 & Max output tokens & 8192 \\
 & Thinking level & \texttt{HIGH} \\
 & Output format & JSON-only; schema-enforced \\
 & Controlled vocabulary & 11 canonical \texttt{process\_label}s \\
\midrule
\textbf{Grounding} 
 & Strict grounding & Enabled (invalid/missing references $\rightarrow$ reject \& retry) \\
 & Citation types (week$\times$dose) & \texttt{obs:\textless{}id\textgreater{}}, \texttt{jump:\textless{}id\textgreater{}}, \texttt{lit:\textless{}id\textgreater{}} \\
 & Citation types (hierarchical) & \texttt{obs:}, \texttt{jump:}, \texttt{lit:}, \texttt{path:\textless{}Reactome\_id\textgreater{}} \\
\midrule
\textbf{Hierarchy} 
 & week$\times$dose $\rightarrow$ dose\_time & Evidence preserved with per-week catalogs \\
 & dose\_time $\rightarrow$ global & Cross-dose summarization with evidence-ID propagation \\
\bottomrule
\end{tabular}
\end{table}

\begin{table}[htbp]
  \centering
  \small
  \caption{Delta and baseline diagnostics.}
  \label{tab:s2_delta_diag}
  \resizebox{\columnwidth}{!}{%
\begin{tabular}{@{} l c c c c @{}}
\toprule
& \multicolumn{2}{c}{\textbf{Control}} & \multicolumn{2}{c}{\textbf{Treated}} \\
\cmidrule(lr){2-3} \cmidrule(l){4-5}
\textbf{Feature} ($\Delta$ z-score) & \textbf{Median} $|\Delta|$ & \textbf{q95} $|\Delta|$ & \textbf{Median} $|\Delta|$ & \textbf{q95} $|\Delta|$ \\
\midrule
\texttt{area\_median}             & 0.520 & 2.738 & 2.080 & 6.484 \\
\texttt{perimeter\_median}        & 0.374 & 2.550 & 2.322 & 6.008 \\
\texttt{eccentricity\_median}     & 0.545 & 1.517 & 1.379 & 4.745 \\
\texttt{solidity\_median}         & 1.041 & 2.383 & 1.399 & 4.758 \\
\texttt{mean\_intensity\_median}  & 0.533 & 1.473 & 1.229 & 3.892 \\
\texttt{glcm\_contrast\_median}   & 0.728 & 2.503 & 1.260 & 4.500 \\
\texttt{glcm\_correlation\_median}& 0.899 & 3.124 & 1.489 & 5.199 \\
\texttt{glcm\_energy\_median}     & 0.425 & 2.736 & 2.164 & 5.933 \\
\texttt{glcm\_homogeneity\_median}& 0.433 & 1.570 & 1.871 & 4.544 \\
\bottomrule
\multicolumn{5}{@{}l@{}}{\footnotesize \textit{Note:} Computed on well-level delta profiles.}\\
\multicolumn{5}{@{}l@{}}{\footnotesize ($\Delta = z - \mathrm{median}(z_{\mathrm{control}},\,\mathrm{week})$).} \\
\multicolumn{5}{@{}l@{}}{\footnotesize $n_{\mathrm{control}}=34$, $n_{\mathrm{treated}}=180$.}
\end{tabular}%
}
\end{table}
\FloatBarrier

\begin{table}[htbp]
  \centering
  \small
  \caption{Process label usage summary.}
  \label{tab:s3_label_usage}
  \resizebox{\columnwidth}{!}{%
\begin{tabular}{@{} l c c c c @{}}
\toprule
\textbf{Process label} & \textbf{week$\times$dose} & \textbf{dose\_time} & \textbf{Global} & \textbf{Global dose} \\
& \textbf{hypotheses} & \textbf{phases} & \textbf{claims} & \textbf{summaries} \\
\midrule
\texttt{dna\_damage\_response}        & 24 (17.6\%) & 1 (5.9\%)  & 1 (25.0\%) & 2 (40.0\%) \\
\texttt{oxidative\_stress}            &  9 (6.6\%)  & 4 (23.5\%) & 1 (25.0\%) & 0 (0.0\%) \\
\texttt{inflammation\_innate\_immune} &  8 (5.9\%)  & 3 (17.6\%) & 1 (25.0\%) & 1 (20.0\%) \\
\texttt{cell\_cycle\_arrest}          &  8 (5.9\%)  & 6 (35.3\%) & 0 (0.0\%)  & 2 (40.0\%) \\
\texttt{senescence\_like}             & 10 (7.4\%)  & 2 (11.8\%) & 1 (25.0\%) & 2 (40.0\%) \\
\texttt{apoptosis\_cell\_death}       &  6 (4.4\%)  & 4 (23.5\%) & 1 (25.0\%) & 2 (40.0\%) \\
\texttt{mitochondrial\_dysfunction}   &  3 (2.2\%)  & 3 (17.6\%) & 1 (25.0\%) & 0 (0.0\%) \\
\texttt{proteostasis\_upr}            & 17 (12.5\%) & 5 (29.4\%) & 1 (25.0\%) & 2 (40.0\%) \\
\texttt{metabolic\_reprogramming}     & 25 (18.4\%) & 8 (47.1\%) & 1 (25.0\%) & 2 (40.0\%) \\
\texttt{adaptation\_recovery}         & 23 (16.9\%) & 4 (23.5\%) & 1 (25.0\%) & 1 (20.0\%) \\
\texttt{other\_uncertain}             &  3 (2.2\%)  & 2 (11.8\%) & 0 (0.0\%)  & 0 (0.0\%) \\
\bottomrule
\multicolumn{5}{@{}l@{}}{\footnotesize \textit{Note:} Denominators: week$\times$dose hypotheses $n=136$; dose\_time phases $n=17$;} \\
\multicolumn{5}{@{}l@{}}{\footnotesize global claims $n=4$; global dose summaries $n=5$.}
\end{tabular}%
}
\end{table}
\FloatBarrier

\begin{table}[htbp]
  \centering
  \small
  \caption{Grounding reference statistics (evidence-reference coverage and counts).}
  \label{tab:s4_grounding}
  \resizebox{\columnwidth}{!}{%
\begin{tabular}{@{} l c l @{}}
\toprule
\textbf{Metric} & \textbf{Value} & \textbf{Notes} \\
\midrule
\multicolumn{3}{@{}l}{\textbf{Run scale}} \\
Week$\times$dose records & 45 & Prompt instances \\
Hypotheses (total) & 136 & Across all records \\
Hypotheses per record (mean) & 3.022 & \\
Hypotheses per record (min--max) & 2--4 & Distribution: \{2:1, 3:42, 4:2\} \\
\midrule
\multicolumn{3}{@{}l}{\textbf{Evidence references per hypothesis}} \\
Quant refs per hypothesis (mean) & 1.000 & \texttt{obs:<id>} citations \\
Quant refs per hypothesis (min--max) & 1--1 & \\
Retrieval refs per hypothesis (mean) & 2.838 & \texttt{jump:}/\texttt{lit:} citations \\
Retrieval refs per hypothesis (min--max) & 1--5 & \\
\midrule
\multicolumn{3}{@{}l}{\textbf{Retrieval evidence usage}} \\
Hypotheses citing JUMP neighbors & 124 (91.2\%) & At least one \texttt{jump:} ref \\
Hypotheses citing literature & 102 (75.0\%) & At least one \texttt{lit:} ref \\
Hypotheses citing both \texttt{jump} + \texttt{lit} & 90 (66.2\%) & \\
Total retrieval refs (\texttt{jump}) & 219 & Count across all hypotheses \\
Total retrieval refs (\texttt{lit}) & 167 & Count across all hypotheses \\
\midrule
\multicolumn{3}{@{}l}{\textbf{Grounding caps (per record)}} \\
Allowed obs IDs & 1 & \\
Allowed JUMP refs & 10 & \\
Allowed literature refs & 10 & \\
\bottomrule
\end{tabular}%
}
\end{table}
\FloatBarrier

\begin{table}[htbp]
  \centering
  \small
  \caption{V2 proxy mapping (process labels to proxy feature sets). }
  \label{tab:s5_proxy_map}
  \begin{tabular}{@{} l p{0.55\columnwidth} @{}}
\toprule
\textbf{Process label} & \textbf{Proxy expectations} \\
& \textbf{(reduced morphology feature set)} \\
\midrule
\texttt{dna\_damage\_response}        & \texttt{glcm\_contrast(+)}, \texttt{glcm\_correlation(0)}, \texttt{mean\_intensity(0)}, \texttt{area(0)} \\
\addlinespace
\texttt{oxidative\_stress}            & \texttt{glcm\_contrast(+)}, \texttt{mean\_intensity(0)}, \texttt{area(0)} \\
\addlinespace
\texttt{inflammation\_innate\_immune} & \texttt{glcm\_contrast(+)}, \texttt{area(0)} \\
\addlinespace
\texttt{cell\_cycle\_arrest}          & \texttt{area(+)}, \texttt{perimeter(+)}, \texttt{eccentricity(0)}, \texttt{solidity(0)} \\
\addlinespace
\texttt{senescence\_like}             & \texttt{area(+)}, \texttt{perimeter(+)}, \texttt{solidity(0)}, \texttt{eccentricity(0)}, \texttt{glcm\_contrast(+)}, \texttt{glcm\_correlation(0)}, \texttt{glcm\_energy(-)}, \texttt{mean\_intensity(0)} \\
\addlinespace
\texttt{apoptosis\_cell\_death}       & \texttt{area(-)}, \texttt{perimeter(-)}, \texttt{mean\_intensity(0)}, \texttt{glcm\_contrast(+)}, \texttt{glcm\_energy(-)} \\
\addlinespace
\texttt{mitochondrial\_dysfunction}   & \texttt{mean\_intensity(0)}, \texttt{glcm\_contrast(+)} \\
\addlinespace
\texttt{proteostasis\_upr}            & \texttt{glcm\_contrast(+)}, \texttt{glcm\_energy(-)} \\
\addlinespace
\texttt{metabolic\_reprogramming}     & \texttt{mean\_intensity(0)}, \texttt{area(0)} \\
\addlinespace
\texttt{adaptation\_recovery}         & \texttt{area(0)}, \texttt{glcm\_contrast(0)} \\
\addlinespace
\texttt{other\_uncertain}             & (none) \\
\bottomrule
\multicolumn{2}{@{}p{0.95\columnwidth}@{}}{\footnotesize \textit{Note:} Direction codes: (+) expected increase; (-) expected decrease; (0) either direction / weak proxy. Feature mappings are derived from canonical morphological hallmarks of cell stress \cite{HernandezSegura2018Senescence, Kroemer2009Apoptosis, Rohban2017Morphological}. Used strictly as a quantitative consistency check (V2), not as a biological ground truth.} \\
\multicolumn{2}{@{}p{0.95\columnwidth}@{}}{\footnotesize Used as a consistency check, not biological ground truth.}
\end{tabular}
\end{table}

\FloatBarrier

\setcounter{figure}{0}
\renewcommand{\thefigure}{A\arabic{figure}}

\begin{figure}[htbp]
  \centering
  \includegraphics[width=\columnwidth]{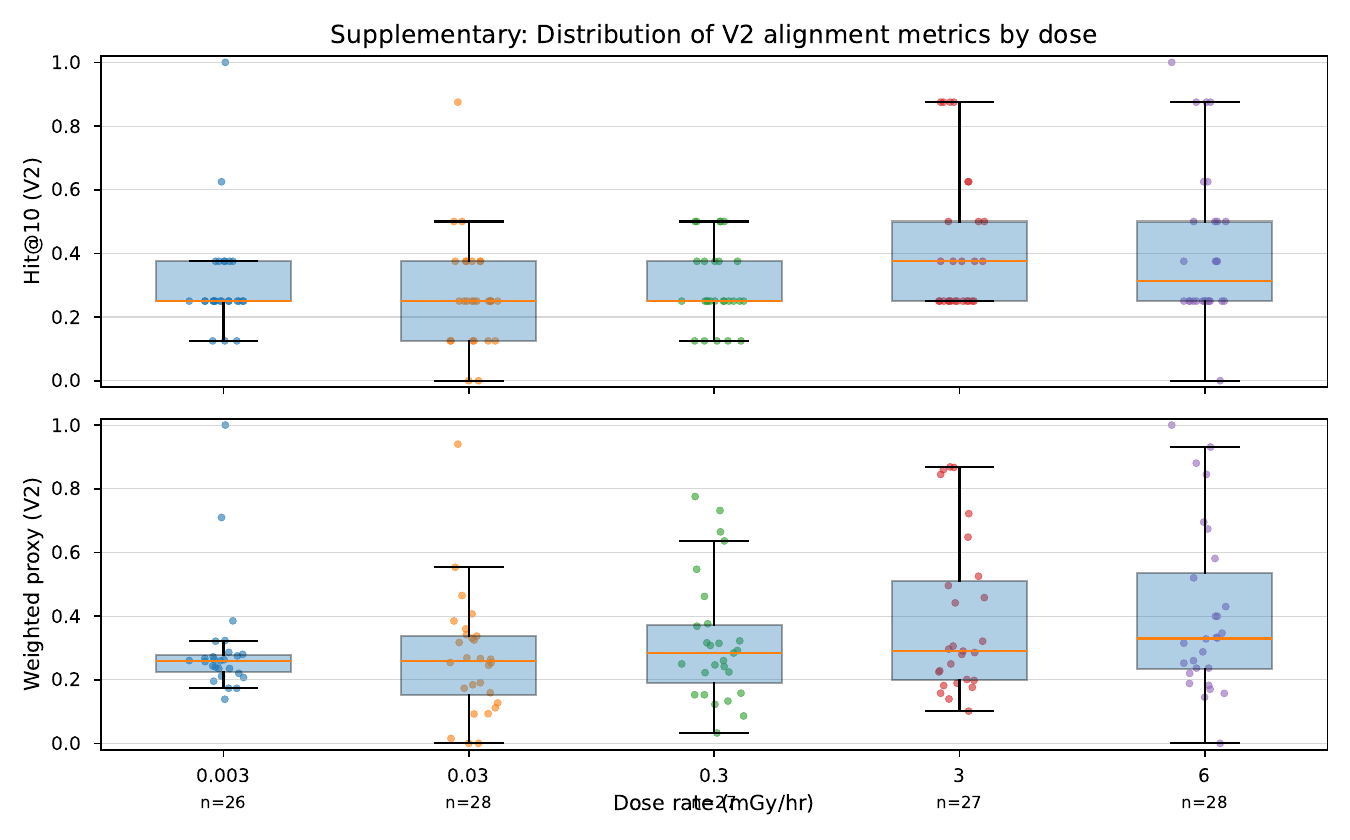}
  \caption{V2 metric distributions.}
  \Description{Histograms and boxplots of V2 Hit@K and weighted proxy scores across reasoning scales.}
  \label{fig:s1_v2_distrib}
\end{figure}

\begin{figure}[htbp]
  \centering
  \includegraphics[width=\columnwidth]{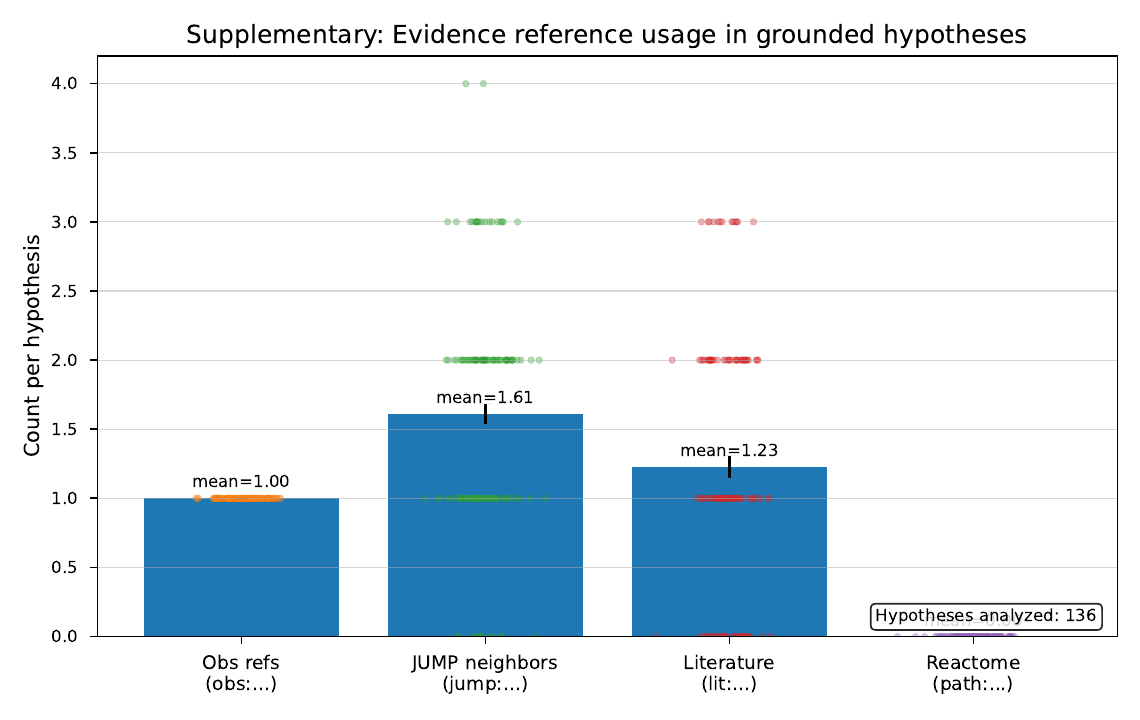}
  \caption{Evidence reference composition.}
  \Description{Bar chart showing counts of observation and retrieval evidence references cited by hypotheses.}
  \label{fig:s2_ref_comp}
\end{figure}

\begin{figure}[htbp]
  \centering
  \includegraphics[width=\columnwidth]{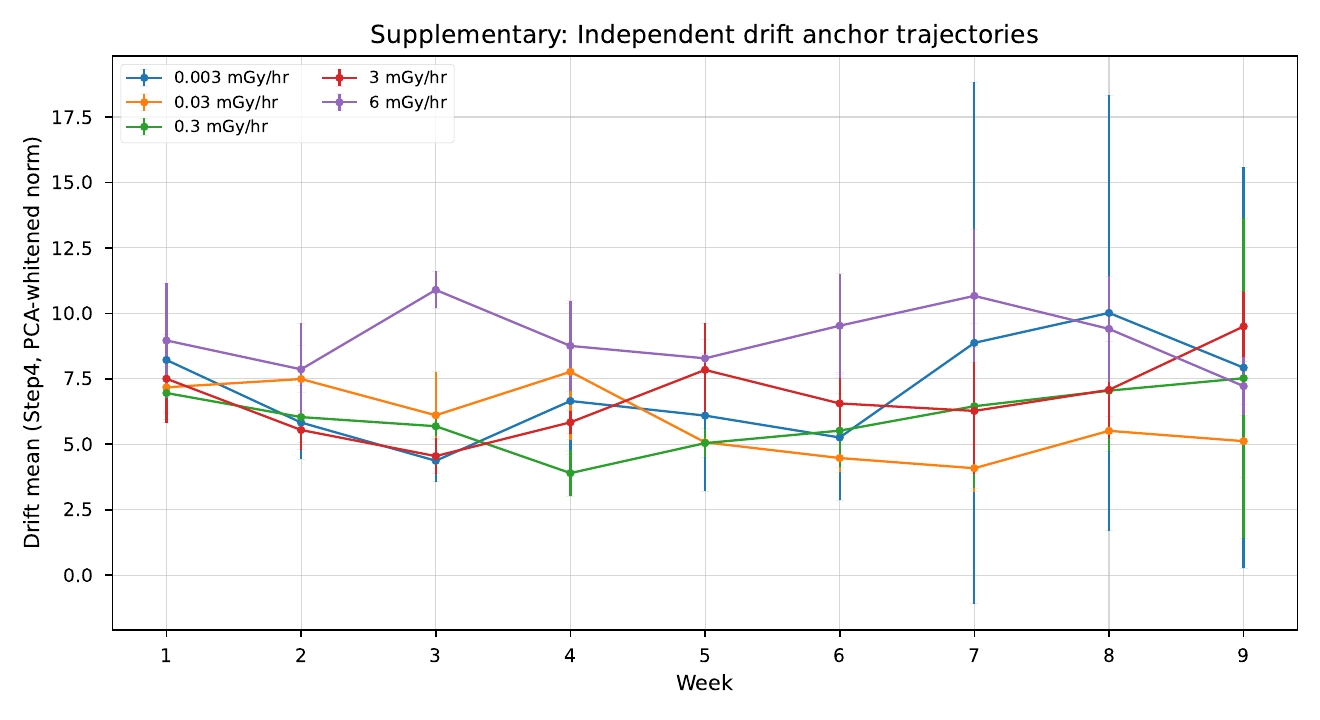}
  \caption{Step~4 drift trajectories.}
  \Description{Line plots of morphology drift magnitude over weeks for each dose rate with uncertainty bands.}
  \label{fig:s3_deltas}
\end{figure}

\begin{figure}[htbp]
  \centering
  \includegraphics[width=\columnwidth]{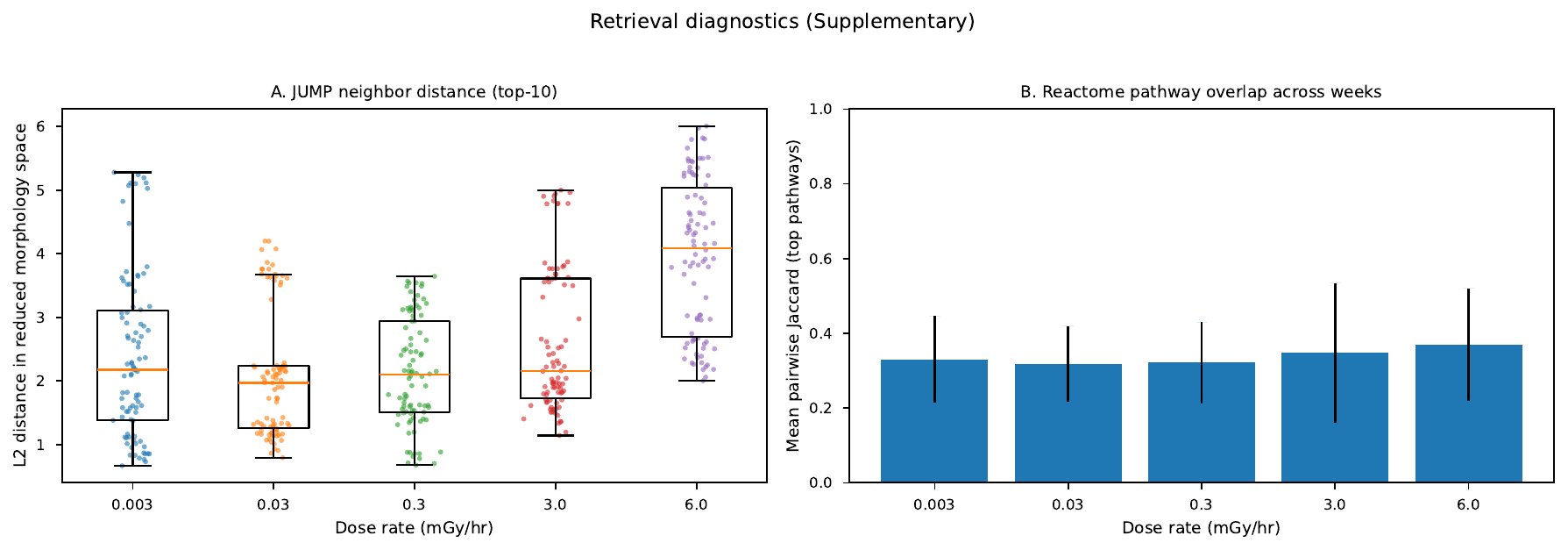}
  \caption{Retrieval diagnostics.}
  \Description{Plots summarizing nearest-neighbor retrieval distance statistics across week-by-dose queries.}
  \label{fig:s4_neighbor_dist}
\end{figure}

\begin{figure}[htbp]
  \centering
  \includegraphics[width=\columnwidth]{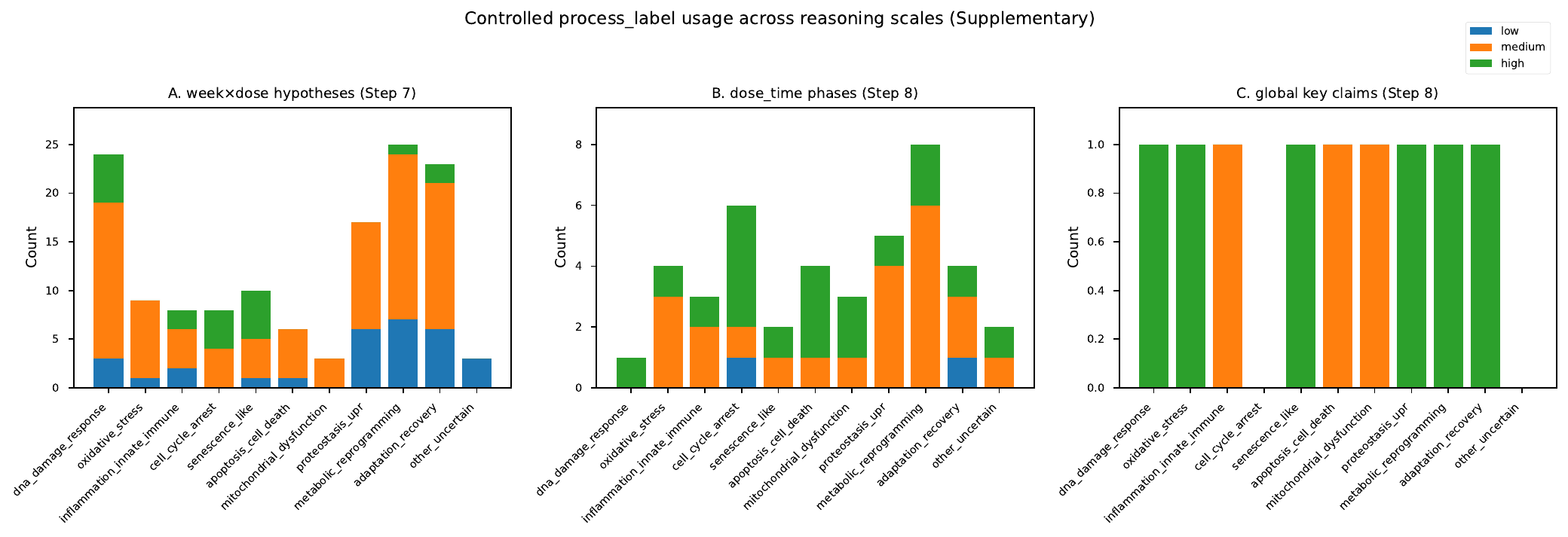}
  \caption{Controlled vocabulary usage across reasoning scales.}
  \Description{Plots comparing process label frequencies across week-by-dose, dose\_time, and global outputs.}
  \label{fig:s5_traj_ci}
\end{figure}

\begin{table}[htbp]
  \centering
  \small
  \renewcommand{\arraystretch}{1.4}
  \setlength{\tabcolsep}{4pt} 
  \caption{Structured Example of a Global-Level Integrated Claim}
  \label{tab:global_claim_example}

  \resizebox{\columnwidth}{!}{%
    \begin{tabular}{|p{0.22\columnwidth}|p{0.74\columnwidth}|}
      \hline
      \textbf{Field} & \textbf{Content} \\ \hline \hline
      \textbf{Claim ID} & claim\_low\_dose\_adaptation \\ \hline
      \textbf{Claim Statement} &
      Chronic low-dose radiation (0.003–0.3 mGy/hr) induces a distinct adaptive phenotype characterized by metabolic reprogramming and proteostatic stress, contrasting with the hypertrophic senescence seen at higher doses. \\ \hline
      \textbf{Process Labels} & metabolic\_reprogramming, adaptation\_recovery, proteostasis\_upr \\ \hline
      \textbf{Quantitative Evidence} & obs:obs\_w6\_d0.003, obs:obs\_w9\_d0.03, obs:obs\_w7\_d0.3 \\ \hline
      \textbf{Retrieval References} & jump:row\_862582, jump:row\_928254, lit:q1:1 \\ \hline
      \textbf{Confidence Level} & High \\ \hline
    \end{tabular}%
  }
\end{table}

\subsection{Coarse Dose-Level Sanity Check}
\label{sec:app_dose_sanity}

\begin{figure}[htbp]
  \centering
  \includegraphics[width=\columnwidth]{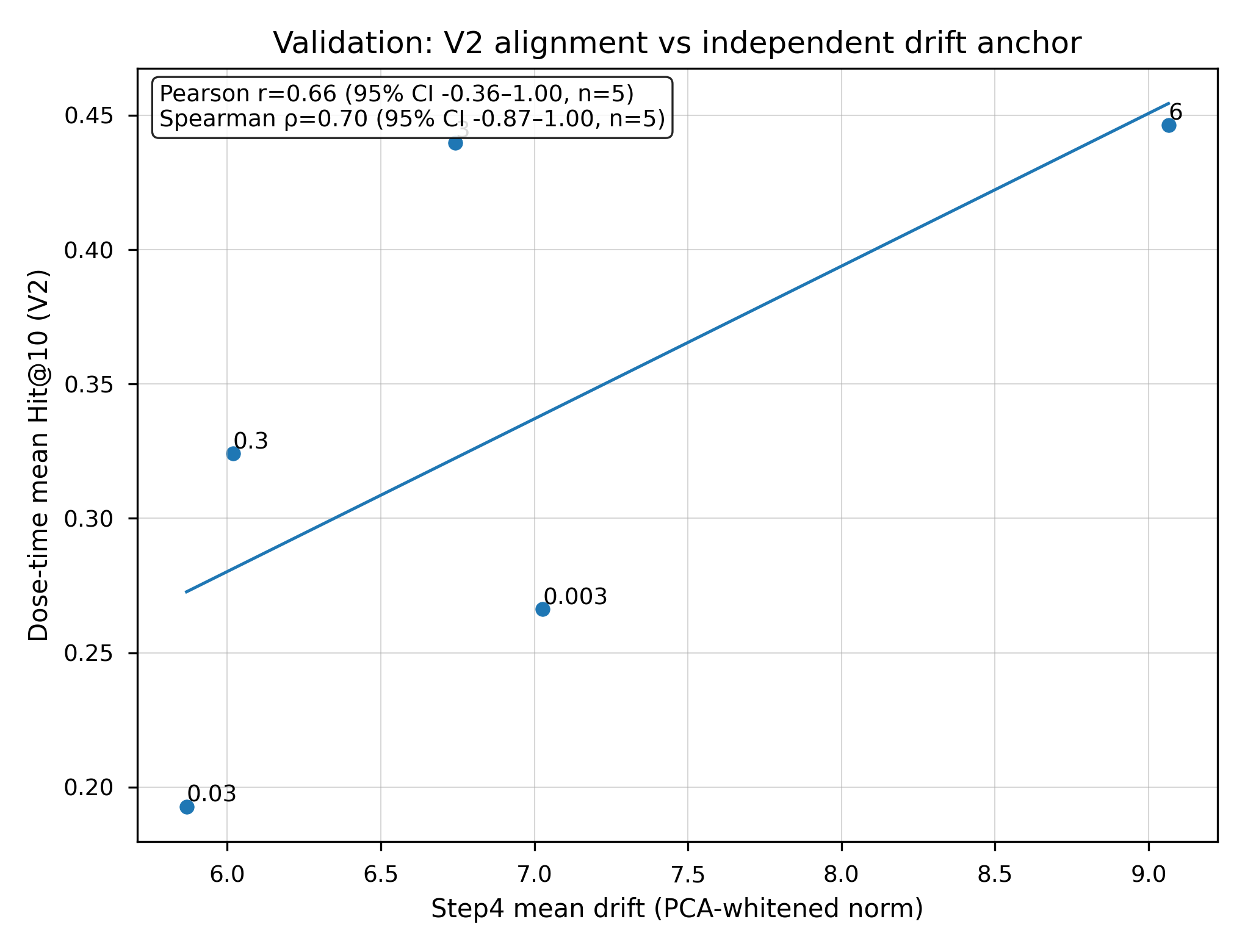}
  \caption{Coarse dose-level comparison ($n=5$) evaluating the consistency between the LLM-external morphology drift and the \texttt{dose\_time} mean Hit@10 summary. While a positive trend is visible, the small sample size limits statistical confidence.}
  \label{fig:app_dose_sanity}
\end{figure}

As a complementary coarse summary to the higher-resolution week$\times$dose analysis ($n=45$) presented in the main text (Section 4.5), we also evaluated a dose-level comparison ($n = 5$ doses). As shown in Figure~\ref{fig:app_dose_sanity}, the association remained positive for the \texttt{dose\_time} mean Hit@10 summary (Pearson $r = 0.661$ with 95\% bootstrap CI $[-0.363, 1.000]$; Spearman $\rho = 0.700$ with 95\% CI $[-0.875, 1.000]$). 

Given the extremely small sample size, the confidence intervals are understandably wide. We retain this five-point analysis strictly as a low-resolution sanity check. Consistent with our approach in the main text, we do not interpret this as an independent biological validation, but rather as a basic verification that the model's scores co-vary positively with the expected morphological drift at the macro-dose level.

\clearpage

\end{document}